\documentclass[12pt]{article}

\usepackage{epsfig}
\usepackage{amsmath}
\usepackage{soul,color}
\usepackage{bm}
\usepackage{epsfig}
\usepackage{natbib}
\usepackage{setspace}
\newcommand{\tr}{^{\prime}}
\def\b#1{\mbox{\boldmath $#1$}}    
\def\bl#1{\mbox{\footnotesize\boldmath {$#1$}}} 

\def\cg#1{\mbox{${\cal #1}$}}      
\def\cgl#1{\mbox{\scriptsize {${\cal #1}$}}}

\newcommand{\ot}{\mbox{$\:\otimes \:$}}
\newcommand{\diag}{{\rm diag}}      
\renewcommand{\th}{\theta}
\newcommand{\Th}{\Theta}

\newcommand{\be}{\beta}
\newcommand{\de}{\delta}
\newcommand{\la}{\lambda}

\newcommand{\ga}{\gamma}

\def\bTheta{\mbox{\boldmath$\Theta$}}
\def\btheta{\mbox{\boldmath$\theta$}}

\def\bxi{\mbox{\boldmath$\xi$}}
\def\bX{\mbox{\boldmath$X$}}
\def\bx{\mbox{\boldmath$x$}}

\def\baselinestretch{1.7}

\usepackage{bm}
\usepackage{epsfig}

\begin{document}
\title{\vspace*{-1.5cm}
A class of Multidimensional Latent Class IRT models for ordinal polytomous item responses}
\author{Silvia Bacci\footnote{Department of Economics, Finance and Statistics,
University of Perugia, Via A. Pascoli, 20, 06123 Perugia.}
\footnote{{\em email}: sbacci@stat.unipg.it} ,
Francesco Bartolucci$^*$\footnote{{\em email}: bart@stat.unipg.it} ,
Michela Gnaldi$^*$\footnote{{\em email}: gnaldi@stat.unipg.it}}  \maketitle
%
\def\baselinestretch{1.3}
\singlespacing
\begin{abstract}\noindent
We propose a class of Item Response Theory models for items with ordinal polytomous
responses, which extends an existing class of multidimensional models for
dichotomously-scored items measuring more than one latent trait.
In the proposed approach,
the random vector used to represent the latent traits is assumed to
have a discrete distribution with support points corresponding to different latent
classes in the population. We also allow for different parameterizations
for the conditional distribution of the response variables given the
latent traits - such as those adopted in the Graded Response model, in the Partial
Credit model, and in the  Rating Scale model - depending on both the type of link
function and the constraints imposed on the
item parameters. For the proposed models we outline how to perform
maximum likelihood estimation via the Expectation-Maximization algorithm.
Moreover, we suggest a strategy for model selection which is based
on a series of steps consisting of selecting specific features, such as the number
of latent dimensions, the number of latent classes, and the specific parametrization.
In order to illustrate the proposed
approach, we analyze data deriving from a study on anxiety and depression as
perceived by oncological patients.

\end{abstract}
\newpage
\section{Introduction}\label{sec:introduction}
Item Response Theory (IRT) models are commonly used to analyze data
deriving from the administration of questionnaires made of items
with dichotomous or polytomous responses (also known, in the
educational setting, as dichotomously or polytomously-scored items).
Dichotomous responses are usually labelled as true or false, right or wrong, yes or no,
whereas polytomous responses correspond to more than two options. Polytomous responses
include both nominal and ordinal responses. In the former, there is no natural ordering
in the item response categories.
In the latter, which are of our interest here,
each item has responses corresponding to
a number of ordered categories (e.g.,
correct, partially correct, wrong).
While nominal polytomous items are especially used to
investigate customers' choices and preferences,
ordinal polytomous items are widespread in several contexts, such as
in education, marketing, and psychology.
For a review about polytomous IRT models,
see \cite{ham:swam:85}, \cite{lin:ham:97}, and
\cite{ner:ost:10}.

A number of models have been proposed in the psychometrical and statistical literature
to analyze items with ordinal polytomous responses,
and several taxonomies can be adopted.
Among the most known, we remind those due to  \cite{sam:72},   \cite{mol:83}, and \cite
{thi:ste:86} which, even though developed independently one another, are strongly
related and overlapping \citep{sam:96, hem:01}.
Combining these parameterizations with possible constraints on item discriminating and
difficulty parameters, the most well
known IRT models for polytomous responses result,
such as the Graded Response model \cite[GRM;][]{sam:69}, the Partial Credit model
\cite[PCM;][]{mast:82}, the Rating Scale model \citep[RSM;][]{andr:78}, and the
Generalized Partial Credit Model \citep[GPCM;][]{mur:92}.
These models are based on the unidimensionality assumption and,
for some of them,
the normality assumption of this latent trait is explicitly introduced.

Several extensions of traditional IRT models for polytomous responses have been proposed in the
literature in order to overcome some
restrictive assumptions and to make the models more flexible and realistic. Firstly,
some authors dealt with multidimensional extensions of IRT models to take into account
that questionnaires are often designed to measure more than one latent trait. Among the
main  contributions in the context of IRT models for polytomous responses, we remind
\cite{dun:ste:87}, \cite{agr:93} and \cite{kel:rij:94}, who proposed a number of
examples of loglinear multidimensional IRT models, \cite{kel:96} for a multidimensional
version of the PCM, and \cite{ada:wil:wan:97} for a wide class of Rasch type
\citep{rasch:60, wri:mas:82} extended  models;
see \cite{rec:09} for a thorough overview of this topic.

Another advance in the IRT literature
concerns the assumption that the population
under study is composed by homogeneous classes of individuals who have very similar
latent characteristics  \citep{laza:henr:68,good:74}.
In some contexts, where the aim is to cluster individuals, this is a convenient
assumption; in health care, for instance, by introducing this assumption
we single out
a certain number of clusters of patients receiving the same clinical treatment.
Secondly, this assumption allows us to estimate the
model in a semi-parametric way, namely
without formulating any assumption on the latent trait distribution. 
Moreover, it is possible to implement the maximum marginal likelihood method making use of the Expectation-Maximization (EM) algorithm \citep{demp:lair:rubi:77}, skipping in this way the problem of intractability of multidimensional integral which characterizes the marginal likelihood when a continuous latent variable is assumed. 
At this regards, \cite{chr:02} outline, through a simulation study, the computational problems encountered during the estimation process of a multidimensional model based on a multivariate normally distributed ability.    
See also \cite{mas:85}, \cite{lan:rost:88}, \cite{hei:96}, and 
\cite{for:07} for a comparison
between traditional IRT models
with those formulated by a latent class approach.
For some examples of discretized variants of IRT models we also
remind 
\cite{lind:91}, \cite{for:92}, \cite{hoi:mol:97}, \cite{ver:01}, and \cite{smit:03}. Another interesting
example of
combination between the IRT approach and 
latent class approach is represented by the mixed Rasch
model for ordinal polytomous data 
 \citep{rost:91, dav:rost:95},  builded as a mixture of latent classes with a separate Rasch model assumed to hold within each of these classes.

As concerns the combination of the two above mentioned extensions,
in the context of dichotomously-scored items \cite{bart:07}
proposed a class of
multidimensional latent class (LC) IRT models, where:
({\em i}) more latent traits are simultaneously considered 
and each item is associated with only one of them
(between-item multidimensionality - for details see \cite{ada:wil:wan:97, zhang:04}) and ({\em ii}) these latent traits are represented by a
random vector with a discrete distribution common to all subjects
(each support point of such a distribution identifies a different
latent class of individuals). Moreover, in this class of models
either a Rasch \citep{rasch:60} or a
two-parameter logistic (2PL) parameterization \citep{birn:68} may be adopted for
the probability of a correct response to each item. 
Similarly to \cite{bart:07}, \cite{dav:08} proposed the diagnostic model, which, as main difference, assumes fixed rather than free abilities. An interesting comparison of multidimensional IRT models based on continuous and discrete latent traits was performed by \cite{hab:dav:lee} in terms of goodness of fit, similarity of parameter estimates and computational time required.  

The aim of the present paper is to extend the class of models of \cite{bart:07}
to the case of items for ordinal polytomous responses. The proposed extension is
formulated so that different
parameterizations may be adopted for the conditional
distribution of the response variables, given the latent traits. We mainly refer to the
classification
criterion proposed by  \cite{mol:83}; see also \cite{agr:90} and \cite{ark:01}.
Relying on the
type of link function, it allows to discern among: ({\em i}) graded response
models, based on global (or cumulative) logits; ({\em ii}) partial credit models, which
make use of
local (or adjacent category) logits; and ({\em iii}) sequential models, based on
continuation ratio logits. For each of these link functions, 
we explicitly consider the possible
presence of constraints on item discrimination parameters and threshold difficulties. As
concerns the first element, we take into account the possibility that all items have the
same discriminating power against the possibility that they discriminate differently.
Moreover, we discern the case in which each item differs from the others for
different distances  between the difficulties of consecutive response categories
and the special case in
which the distance between difficulty levels from category to category
is the same for
all items. On the basis of the choice of all the mentioned features
(i.e., type of link function, item discriminant parameters, item difficulties),
different
parameterizations for ordinal responses are defined. We show how these parameterizations
result in an extension of traditional IRT models, by introducing assumptions of
multidimensionality and discreteness of latent traits.

In order to estimate each model in the proposed class,
we outline an EM algorithm.
Moreover, special attention is given to the model selection procedure, that aims 
at choosing the optimal number of latent classes, the type of link function, the
number of latent dimensions and the
allocation of items within each dimension, and the
parameterization for the item discriminating and difficulty parameters.

In order to illustrate the proposed class of models, we analyze
a dataset collected by a questionnaire on anxiety and depression of
oncological patients, and formulated following the ``Hospital
Anxiety and Depression Scale'' (HADS) developed by \cite{zig:sna:1983}. Through this
application, each step of the model selection procedure is illustrated and the
characteristics of each latent class, in terms of estimated levels of the latent traits,
are described with reference to the selected model.

In summary, the proposed class of models allows for ({\em i}) ordinal polytomous responses of different nature, ({\em ii}) multidimensionality  and ({\em iii})  discreteness of latent traits, at the same time.  As concerns the first point, our model includes different link functions that are suitable for a wide type of empirical data. Moreover, our formulation allows for estimating both abilities and probabilities, and the introduction of latent classes represents a semi-parametric approach that computationally simplifies, through an EM algorithm,  the maximization of log-likelihood function during the estimation process. To our knowledge, there are not other contributions treating all these topics in a same unifying framework, even if the single aspects are separately included in several existing types of models, as above outlined.

The reminder of this paper is organized as follows. In Section \ref{prelim} we describe
some basic parameterizations
for IRT models for items with ordinal responses.  In Section
\ref{poly} we describe the proposed class of multidimensional LC IRT models for items
with ordinal responses. Section \ref{sec:inference} is devoted to
maximum likelihood estimation which is
implemented through an EM algorithm; moreover, in the same section
we treat the issue of model selection.
In Section \ref{applic}, the proposed class of  models is illustrated through
the analysis of a real dataset,
whereas some final remarks are reported in Section \ref{sec:conclusion}.
\section{Models for polytomous item responses}
\label{prelim}
Let $X_j$ denote the response variable for the $j$-th item of the
questionnaire, with $j=1,\ldots,r$. This variable has $l_j$ categories, indexed
from 0 to $l_j-1$. Moreover, in the unidimensional case,
let
\[
\la_{jx}(\th)=p(X_j=x|\Th=\th),\quad x=0,\ldots,l_j-1,
\]
denote the probability that a subject with latent trait
(or ability) level $\th$ responds by category
$x$ to this item.
Also let $\b\la_j(\th)$ denote the probability vector
$(\la_{j0}(\th),\ldots,\la_{j,l_j-1}(\th))\tr$,
the elements of which sum up to 1.

The IRT models for polytomous responses that are here of interest may be expressed
through the general formulation
\begin{equation}\label{eq:poly_gen}
g_x[\b\la_j(\th)] = \ga_j(\theta-\be_{jx}),\quad j=1,\ldots,r,\:
x=1,\ldots,l_j-1,
\end{equation}
where $g_x(\cdot)$ is a link function specific of category $x$ and $\ga_j$ and
$\be_{jx}$ are item parameters which are usually identified as {\em discrimination
indices} and {\em difficulty levels}
and on which suitably constraints may be assumed.

On the basis of the specification of the link function in (\ref{eq:poly_gen}) and on
the basis of the adopted constraint on the item parameters, different unidimensional IRT
models for polytomous responses result. In particular, the formulation of each of these
models depends on:
\begin{enumerate}
\item {\em Type of link function}:
We consider the link based on: ({\em i}) global (or
cumulative) logits; ({\em ii}) local (or adjacent categories) logits;
and ({\em iii}) continuation ratio logits.
In the first case, the link function is defined as
\[
g_x[\b\la_j(\b\th)]
=\log\frac{\la_{x|\th}^{(j)}+\cdots+\la_{l_j-1|\th}^{(j)}}
{\la_{0|\th}^{(j)}+\cdots+\la_{x-1|\th}^{(j)}}
=\log\frac{p(X_{j}\geq x|\theta)}{p(X_{j} < x|\theta)},\quad x=1, \ldots, l_j-1,
\]
and compares the probability that item response is in category $x$ or higher with
the probability that it is in a lower category. Moreover, with local
logits we have that
\[
g_x[\b\la_j(\th)]
=\log\frac{\la_{x|\th}^{(j)}}{\la_{x-1|\th}^{(j)}}
=\log\frac{p(X_{j}=x |\theta)}{p(X_{j}=x-1|\theta)},
\quad x=1,\ldots,l_j-1,
\]
and then the probability of each category $x$ is compared with the probability of the
previous category. Finally, with continuation ratio logits we have that
\[
g_x[\b\la_j(\th)]
=\log\frac{\la_{x+1|\th}^{(j)}+\cdots+\la_{l_j-1|\th}^{(j)}}
{\la_{x|\th}^{(j)}}
=\log\frac{p(X_{j} > x |\theta)}{p(X_{j} = x |\theta)},
\quad x=1,\ldots,l_j-1,
\]
and then the probability of a
response in category $x$ is compared with the probability associated
to the previous category or higher.

Global logits are typically used when the trait of interest is assumed to be continuous
but latent, so that it can be observed only when each subject reaches a given threshold
on the latent continuum. On the contrary, local logits are used to identify one or more
intermediate levels of performance on an item and to award a partial credit for reaching
such intermediate levels. Finally, continuation ratio logit is useful when  sequential
cognitive processes are involved (e.g., problem solving or repeated trials),
how it typically happens in the educational context.
Note that the interpretation of continuation ratio logits
is very different from that of
local logits. The latter ones describe the transition from one category to an adjacent
one given that one of these two categories have been chosen. Thus, each
of these logits excludes any other categories. Differently,
continuation ratio logits describe the transition between
adjacent categories, given that the smallest between the two has been reached.
IRT models based on global logits are also known as graded response models, those based
on local logits are known as partial credit models. Moreover, IRT models based on
continuation ratio logits are also called sequential models.
\item {\em Constraints on the discrimination parameters}: We consider: ({\em i})
a general
situation in which each item may discriminate differently from the others and 
({\em ii}) a special case in which all the items discriminate in the same way,
that is
\begin{equation}
\gamma_j=1,\quad j = 1, \ldots, r.\label{eq:same_discrimination}
\end{equation}
Note that, in both cases, we assume that,
within each item, all response categories share the same $\gamma_j$, in order to keep
the conditional probabilities away
from crossing and so avoiding degenerate
conditional response probabilities.
\item {\em Formulation of item difficulty parameters}:
We consider:
({\em i}) a general situation in which the parameters $\be_{jx}$ are unconstrained
and ({\em ii}) a special case in which these parameters are constrained so that
the distance between difficulty levels from category to category is the same for
each item (rating scale parameterization). Obviously, the second case
makes sense when all items have the same number of response categories,
that is $l_j=l$, $j=1,\ldots,r$. This constraint may be expressed as
\begin{equation}
\beta_{jx} = \beta_j + \tau_x,\quad j=1,\ldots,r,\: x = 0,\ldots,l-1,
\label{eq:rating_scale}
\end{equation}
where $\beta_j$ indicates the difficulty of item $j$ and $\tau_x$ is the difficulty of
response category $x$ for all $j$.
\end{enumerate}

By combining the above constraints, we obtain four different
specifications of the item parametrization, based on free or constrained discrimination
parameters and on rating scale or free parameterization for difficulties.
Therefore, also according to the type of link function, twelve different types of
unidimensional IRT model for ordinal responses result. These models are
listed in Table \ref{tab:classif}.

\begin{table}[!ht]\centering\vspace*{0.5cm}
{\small
\begin{tabular}{cccccc}
\hline \hline
discrimination     &  difficulty &  resulting  &  \multicolumn3c{resulting model
(depending on the type of logit)} \\\cline{4-6}
indices & levels  & parameterization           &   global   &  local  &
 continuation \\
\hline
          free     &        free       & $\gamma_j (\theta - \beta_{jx})$
          &     GRM                      &    GPCM     &  SM  \\
          free     &   constrained &    $\gamma_j [\theta - (\beta_{j} + \tau_x)]$
          &     RS-GRM              &    RS-RSM    &   RS-SM  \\
        constrained   &   free     &      $ \theta - \beta_{jx}$
        &     1P-GRM               &    PCM   &   SRM          \\
         constrained   &   constrained  &  $ \theta - (\beta_{j} + \tau_x)$
         &     1P-RS-GRM       &    RSM  &   SRSM   \\
\hline
\end{tabular}}
\caption{\em List of unidimensional IRT models for ordinal polytomous responses which result
from the different choices of the link function, constraints on the discrimination
indices, and constraints on the difficulty levels.} \label{tab:classif}
\vspace*{0.5cm}
\end{table}

Abbreviations used for the models specified in Table \ref{tab:classif} refer to the way
the corresponding models are known in the literature. Thus, other than GRM, RSM, PCM,
and GPCM already mentioned in Section \ref{sec:introduction}
it is possible to identify: SM indicating
the Sequential Model obtained as special case of the acceleration model of
\cite{sam:95}, where the acceleration step parameter is constrained to one and the
discriminant
indices are all constant over the response categories;  RS-GRM indicating
the rating scale
version of the GRM introduced by \cite{mur:90}; RS-GPCM  and RS-SM  that are rating
scale versions of GPCM \citep{mur:97} and SM,
respectively;  1P-GRM \citep{ark:01}, 1P-RS-GRM \citep{ark:01}, and SRM \citep
[Sequential Rasch Model;][]{tutz:90} indicating versions with
constant discrimination index corresponding to the GRM, RS-GRM, and SM models,
respectively. Finally, by SRSM
we indicate the Sequential Rating Scale Model of \cite{tutz:90}.
We observe that
Table \ref{tab:classif} identifies a hierarchy of models in correspondence
with each type of link function.

As an illustration, consider that if we choose a global logit link
function and the least restrictive parameterization for the item parameters, we obtain
the GRM, that represents one of the most well known generalization of the 2PL model to
items with ordinal responses. This generalization is based on the assumption
\begin{equation}\label{eq:GRM}
\log \frac{p(X_{j}\geq x|\theta)}{p(X_{j} < x|\theta)}=\gamma_j(\theta - \beta_{jx}),
\quad j=1,\ldots,r,\:x=1, \ldots, l_j-1.
\end{equation}
Moreover, by combining the local logit link and the most restrictive parameterization
for the item parameters, the RSM results.
It represents an extension of the Rasch
model to items with ordinal responses, which is based on the assumption
\begin{equation}\label{eq:RSM}
\log \frac{p(X_{j}=x |\theta)}{p(X_{j}=x-1|\theta)} = \theta - (\beta_{j} + \tau_x),
\quad j=1,\ldots,r,\:x=1, \ldots, l-1.
\end{equation}

Since all the models presented in Table \ref{tab:classif} can be expressed in terms of
nonlinear mixed models \citep{rijmen:03}, a suitable and very common parameter
estimation method  is the maximum marginal log-likelihood (MML) method,
which is based on integrating out
the unknown individual parameters, so that only the item parameters need to be
estimated. To treat the integral characterizing the marginal log-likelihood
function, different approaches can be adopted (\cite{rijmen:03} for details).
Under the assumption that the latent trait has a normal distribution, the
Gauss-Hermite quadrature can be adopted to compute this integral which
is then maximized by a direct method (e.g., Newton-Raphson algorithm) or indirect
(e.g., EM algorithm). Alternatively, we can adopt a quasi-likelihood approach or
a Bayesian approach based on Markov Chain Monte Carlo methods. Once the
model parameters have been estimated, person parameters
can be estimated by treating item parameters as known and maximizing the log-likelihood
with respect to the latent trait or, alternatively, using the
expected value or the maximum value of the corresponding posterior distribution.

Among the above mentioned models, those based on Rasch type parametrization
(i.e., PCM and RSM) may be also estimated through the
conditional maximum likelihood \citep[CML;][]{wri:mas:82} method.
This method allows us to estimate the item parameters without formulating
any assumption on the latent trait distribution. It is based on maximizing the
log-likelihood conditioned on the
individual raw scores that, in the case of
Rasch type models, represent a sufficient statistics for ability parameters.
The resulting function only depends on the difficulty
parameters that, therefore, can be
consistently estimated. \cite{tutz:90} proposed a modified version of CML method to
estimate the SRM and the SRSM. Another estimation method used is the joint or
unconditional maximum likelihood \citep[][for details]{wri:mas:82} which, however,
does not provide consistent parameter estimates.
\section{The proposed class of models}\label{poly}
In the following, we describe the
multidimensional extension of the unidimensional IRT models for ordinal
responses mentioned in the previous section, which is based
on latent traits with a discrete distribution.
We first present the assumptions on which the proposed class of models is based
and, then, a formulation in matrix notation which is useful for the estimation.

We recall that the proposed class of models also represents a generalization to the
case of ordinal polytomous responses of the class of multidimensional models proposed by
\cite{bart:07} for dichotomously-scored items.
\subsection{Basic assumptions}
Let $s$ be the number of different latent traits measured by the items,
let $\bTheta = (\Theta_1,\ldots, \Theta_{s})\tr$ be a vector of latent variables
corresponding to these latent traits, and let $\btheta=
(\theta_1,\ldots,\theta_s)\tr$ denote one of its possible realizations. The random vector $
\bTheta$ is assumed to have a discrete distribution with $k$ support points, denoted by
$\bxi_1,\ldots,\bxi_k$, and probabilities $\pi_1,\ldots,\pi_k$, with $\pi_c = p(\bTheta=
\bxi_c)$.
Moreover, let $\delta_{jd}$ be a dummy variable equal to $1$ if item $j$
measures latent trait of type $d$ and to 0 otherwise, with
$j=1,\ldots,r$ and $d=1, \ldots, s$.

Coherently with the  introduction of  vector $\bTheta$, we redefine the
conditional response probabilities
\[
\la_{jx}(\b\th)=p(X_j=x|\b\Th=\b\th),\quad x=0,\ldots,l_j-1,
\]
and we let $\b\la_j(\b\th)=(\la_{j0}(\b\th),\ldots,\la_{j,l_j-1}(\b\th))\tr$.
Then, assumption (\ref{eq:poly_gen}) is generalized as follows
\begin{equation}\label{eq:poly_gen2}
g_x(\b\la_j(\b\th)) = \ga_j (\sum_{d=1}^{s} \delta_{jd} \theta_d - \beta_{jx}),
\quad j=1,\ldots,r,\:
x=1,\ldots,l_j-1,
\end{equation}
where the item parameters $\ga_j$ and $\be_{jx}$ may be subjected to the same
parametrizations illustrated in Section \ref{prelim}.
More precisely, on the basis of the constraints assumed on these
parameters, we obtain different
specifications of equation (\ref{eq:poly_gen}) which are reported in Table
\ref{tab:classif}, where we distinguish the case of $s=1$ from that of $s>1$.

\begin{table}[!hb]\centering\vspace*{0.5cm}
{\small
\begin{tabular}{ccll}
\hline \hline
discrimination    &  difficulty & \multicolumn2c{Number of latent traits} \\\cline{3-4}
indices & levels  &         \multicolumn1c{$s = 1$}              &
\multicolumn1c{$s > 1$}            \\\hline
    free     &        free       &  $\gamma_j (\theta - \beta_{jx})$
    &     $\gamma_j (\sum_d \delta_{jd} \theta_d - \beta_{jx})$\\
    free     &   constrained & $\gamma_j [\theta - (\beta_{j} + \tau_x)]$
    &     $\gamma_j [\sum_d \delta_{jd} \theta_d - (\beta_{j} + \tau_x)]$\\
    constrained   &   free     &  $ \theta - \beta_{jx}$
    &      $ \sum_d \delta_{jd} \theta_d - \beta_{jx}$  \\
    constrained   &   constrained  & $ \theta - (\beta_{j} + \tau_x)$
    &      $    \sum_d \delta_{jd} \theta_d - (\beta_{j} + \tau_x)$  \\\hline
\end{tabular}}
\caption{\em Resulting item parameterizations for $s=1$ and $s>1$.}  \label{tab:multi_h}
\vspace*{0.5cm}\end{table}

Each of the item parameterizations shown in Table \ref{tab:multi_h}  may be
indifferently combined both with global, local, and continuation ratio logit link functions
to obtain
different types of multidimensional LC
IRT models for ordinal responses, representing as many as generalizations of models as
in Table \ref{tab:classif}. For
instance, we may define the multidimensional LC versions of
GRM, defined through equation (\ref{eq:GRM}), and RSM,
defined through equation (\ref{eq:RSM}), respectively as
\begin{equation}\label{eq:multi_LC_GRM}
\log \frac{p(X_{j}\geq x|\bTheta=\btheta)}{p(X_{j} < x|\bTheta = \btheta)}=
\gamma_j (\sum_{d=1}^{s} \delta_{jd} \theta_d - \beta_{jx}),\quad x=1, \ldots, l_j-1,
\end{equation}
and
\begin{equation}\label{eq:multi_LC_RSM}
\log \frac{p(X_{j}=x |   \bTheta = \btheta)}{p(X_{j}=x-1|   \bTheta = \btheta)}  =
\sum_{d=1}^{s} \delta_{jd} \theta_d - (\beta_{j} + \tau_{x}), \quad x=1,
\ldots, l-1.
\end{equation}
Note that when $l_j=2$, $j=1,\ldots,r$, so that item responses
are binary, equations (\ref{eq:multi_LC_GRM}) and (\ref{eq:multi_LC_RSM}) specialize,
respectively, in the multidimensional LC 2PL model and in the multidimensional LC Rasch
model, both of them described by \cite{bart:07}.

In all cases, the discreteness of the distribution of the random vector $\bTheta$
implies that the manifest distribution of $\b X=(X_1,\ldots,X_r)\tr$
for all subjects in the $c$-th latent class is equal to
\begin{equation}
p(\b x)=p(\bX = \bx) = \sum_{c=1}^{k} p(\bX = \bx|\bTheta=\bxi_c) \pi_c,
\label{eq:prob_mani}
\end{equation}
where, due to the 
classical assumption of {\em local independence}, we have
\begin{eqnarray}
p(\b x|c)=
p(\bX = \bx|\bTheta=\bxi_c) & = & \prod_{j=1}^{r} p(X_{j} = x_j|\bTheta=\bxi_c) =
\nonumber\\
                            & = & \prod_{d=1}^{s} \: \prod_{j\in\cgl J_d}
                            p(X_j = x_j|\Theta_d=\xi_{cd}), 
                            \label{eq:prob_cond}
\end{eqnarray}
where $\cg J_d$ denotes the subset of $\cg J=\{1,\ldots,r\}$ containing the indices
of the items measuring the $d$-th latent trait, with $d=1,\ldots,s$
and $\xi_{cd}$ denoting the $d$-th elements of $\b\xi_c$.

In order to ensure the identifiability of the proposed models, 
suitable constraints on the 
parameters are required. With reference to the general equation (\ref{eq:poly_gen2}),
we  require that, for each latent trait, one discriminant index  is equal to 1 and one
difficulty  parameter  is equal to 0.
More precisely, let $j_d$ be a specific element of $\cg J_d$, say the first.
Then, when the discrimination indices are not constrained to be constant
as in (\ref{eq:same_discrimination}), we assume that
\[
\ga_{j_d} = 1,\quad d=1,\ldots,s.
\]
Moreover, with free item difficulties we assume that
\begin{equation}
\be_{j_d1} = 0,\quad d=1,\ldots,s, 
\label{eq:constraing_beta1}
\end{equation}
whereas with a rating scale parameterization based on (\ref{eq:rating_scale}),
we assume
\begin{equation}
\be_{j_d} = 0,\quad d=1,\ldots,s,\quad\mbox{and}\quad\tau_1=0. 
\label{eq:constraing_beta2}
\end{equation}

Coherently with the mentioned identifiability constraints, the number of free parameters
of a multidimensional LC IRT model with ordinal responses is obtained by summing the
number of free probabilities $\pi_c$, the number of ability
parameters $\xi_{cd}$, the number of free item difficulty parameters $\beta_{jx}$, and
that of free item discrimination parameters $\ga_j$.
We note that the number of free parameters does not depend on the
type of logit, but only on the type of parametrization assumed 
on item discrimination and
difficulty parameters, as shown in Table \ref{tab:num_par}. In any case, the number of
probabilities is equal to $k-1$ and the number of ability
parameters is equal to $sk$. However, the number of free item difficulty parameters
is given by $[\sum_{j=1}^r(l_j-1)-s]$ under an unconstrained difficulties
parameterization and it is given by $[(r-s)+(l-2)]$ under a rating scale
parameterization.
Finally, the number of free item discrimination parameters is equal
to $(r-s)$ under an unconstrained discrimination parameterization, being 0 otherwise.

\begin{table}[!ht]\centering\vspace*{0.5cm}
{\small
\begin{tabular}{cc|l}
\hline \hline
discrimination    &  difficulty & \multicolumn1c{Number of free parameters} \\
indices & levels  &          \multicolumn1c{$(\#{\rm par})$}              \\\hline
    free     &        free       &  $(k-1) + sk + \big[\sum_{j=1}^r(l_j-1)-s\big] + (r-s)$                          \\
    free     &   constrained & $(k-1) + sk + [(r-s) + (l-2)] + (r-s)$         \\
        constrained   &   free     &  $(k-1) + sk + \big[\sum_{j=1}^r(l_j-1)-s\big]$                                             \\
    constrained   &   constrained  & $(k-1) + sk + [(r-s) + (l-2)]$                                 \\
\hline
\end{tabular}}
\caption{\em Number of free parameters for different constraints on item discrimination and difficulty parameters.}  \label{tab:num_par}\vspace*{0.5cm}
\end{table}

\subsection{Formulation in matrix notation}
In order to efficiently implement parameter estimation,
in this section we express the above described class of
models by using the matrix notation. In order to simplify the description,
we consider the case in which every item has the same number of response
categories, that is $l_j=l$, $j=1,\ldots,r$;
the extension to the general case
in which items may also have a different number of response categories is
straightforward. In the following, by $\b 0_a$ we denote
a column vector of $a$ zeros, by $\b O_{ab}$ an $a\times b$ matrix of zeros, by
$\b I_a$ an identity matrix of size $a$, by $\b 1_a$ a column vector of $a$ ones.
Moreover, we use the symbol $\b u_{ab}$ to denote a column vector of $a$ zeros with
the $b$-th element equal to one and
$\b T_a$ to denote an $a\times a$ lower triangular matrix of ones.
Finally, by $\ot$ we indicate the Kronecker product.

As concerns the link function used in (\ref{eq:poly_gen2}), it may be expressed in a
general way to include different types of parameterizations
\citep{glon:mccu:95, colo:forc:01} as follows:
\begin{equation}\label{eq:logits}
\b g[\b\la_j(\b\th)] = \b C\log[\b M\b\la_j(\b\th)],
\end{equation}
where the vector $\b g[\b\la_j(\b\th)]$ has elements
$g_x[\b\la_j(\b\th)]$ for $x=1,\ldots,l-1$. Moreover, $\b C$ is a matrix
of constraints of the type
\[
\b C = (\begin{matrix}-\b I_{l-1}
& \b I_{l-1}\end{matrix}),
\]
whereas, for the global logit link, matrix $\b M$ is equal to
\[
\b M = \begin{array}{ll}
\left(\begin{matrix}
\b T_{l-1} & \b 0_{l-1} \\
\b 0_{l-1} & \b T_{l-1}\tr
\end{matrix}\right),
\end{array}
\]
for the local logit link it is equal to
\[
\b M = \begin{array}{ll}
\left(\begin{matrix}
\b I_{l-1} & \b 0_{l-1} \\
\b 0_{l-1} & \b I_{l-1}
\end{matrix}\right),
\end{array}
\]
and for the continuation ratio logit link it is given by
\[
\b M = \begin{array}{ll}
\left(\begin{matrix}
\b I_{l-1} & \b 0_{l-1} \\
\b 0_{l-1} & \b T_{l-1}\tr
\end{matrix}\right).
\end{array}
\]
How to obtain the probability vector $\b\la_j(\b\th)$ on the basis of a vector
of logits defined as in (\ref{eq:logits}) is described in \cite{colo:forc:01},
where a method to compute the derivative of a suitable vector of canonical
parameters for $\b\la_j(\b\th)$ with respect to these logits may be found.

Once the ability and difficulty parameters are included in the single vector
$\b\phi$ and taking into account that the distribution of $\b\Th$ has $k$ support
points, assumption (\ref{eq:poly_gen2}) may be expressed through the general formula
$$
\b g[\b\la_j(\b\xi_c)] =
\ga_j\b Z_{cj}\b\phi,\quad c=1,\ldots,k,\:j=1,\ldots,r,
$$
where $\b Z_{cj}$ is a suitable design matrix. The structure of the
parameter vector $\b\psi$ and of these design matrices depend on the type of
constraint assumed on the difficulty parameters, as we explain below.

When the difficulty parameters are unconstrained, $\b\phi$ is a
column vector of size $sk+r(l-1)-s$, which is obtained from
$$
(\xi_{11},\ldots,\xi_{1s},\ldots,\xi_{ks},\beta_{11},\ldots,
\beta_{1,l_1-1}, \ldots, \beta_{r,l_r-1})\tr
$$
by removing the parameters constrained to be 0; see (\ref{eq:constraing_beta1}).
Accordingly, for $c=1,\ldots,k$ and $j=1,\ldots,r$, the design matrix
$\b Z_{cj}$ is obtained by removing suitable columns from the matrix
\[
\begin{pmatrix}
\b 1_{l-1}(\b u_{kc}\ot\b u_{sd})\tr & \b u_{rj}\tr\ot\b I_{l-1}
\end{pmatrix},
\]
where $d$ is the dimension measured by item $j$.
On the other hand, under a rating scale parameterization, $\b\phi$ is a vector of
size $sk+(r-s)+(l-2)$ which is obtained from
$$
(\xi_{11},\ldots,\xi_{1s},\ldots,\xi_{ks},\beta_1,
\ldots,\beta_r,\tau_1,\ldots,\tau_{l-1})\tr
$$
by removing the parameters constrained to be 0 in (\ref{eq:constraing_beta2}).
Accordingly, the design matrix $\b Z_{cj}$ is obtained by removing
specific columns from
\[
\begin{pmatrix}
\b 1_{l-1}(\b u_{kc}\ot\b u_{sd})\tr & \b 1_{l-1}\b u_{rj}\tr & \b I_{l-1}
\end{pmatrix},
\]
where, again, $d$ is the dimension measured by item $j$.
\section{Likelihood inference} \label{sec:inference}
In this section, we deal with likelihood inference for the models proposed in the
previous section. In particular, we first show how to compute the model log-likelihood
and how to maximize it by an EM algorithm. Finally, we deal with model selection. All the computational procedures are implemented in Matlab and R and are available on request from authors.

\subsection{Model estimation}
On the basis of an observed sample of dimension $n$,
the log-likelihood of a model formulated as proposed in Section
\ref{poly} may be expressed as
\[
\ell(\b\eta)=\sum_{\bl x} n_{\bl x}\log[p(\b x)],
\]
where $\b\eta$ is the vector containing all the free model parameters,
$n_{\bl x}$ is the frequency of the response configuration $\b x$,
$p(\b x)$ is computed according to (\ref{eq:prob_mani}) and (\ref{eq:prob_cond})
as a function of $\b\eta$, and by $\sum_{\bl x}$ we mean
the sum extended to all the possible response configurations $\b x$.

In oder to maximize $\ell(\b\eta)$ with respect to $\b\eta$ we
use an EM algorithm \citep{demp:lair:rubi:77} that is implemented
in a similar way as described in \cite{bart:07}, to which we refer
for some details. First of all, denoting by $m_{c,\bl x}$ the (unobserved)
frequency of the response configuration $\b x$ and the latent
class $c$, the {\em complete} log-likelihood is equal to
\begin{equation}
\ell^*(\b\eta)=\sum_c\sum_{\bl x}m_{c,\bl x}\log[p(\b x|c)\pi_c].
\label{eq:comp_lk}
\end{equation}
Now we denote by $\b\eta_1$ the subvector of $\b\eta$
which contains the free latent class probabilities
and by $\b\eta_2$ the subvector containing the remaining free parameters.
More precisely, we let $\b\eta_1=\b\pi$, with $\b\pi=(\pi_2,\ldots,\pi_k)\tr$,
and $\b\eta_2=(\b\ga\tr,\b\phi\tr)\tr$, where $\b\ga$ is obtained by removing
from $(\ga_1,\ldots,\ga_r)\tr$ the parameters which are constrained to be
equal to 1 to ensure identifiability.
Obviously, $\b\ga$ is not present when constraint
(\ref{eq:same_discrimination}) is adopted. Then, we can decompose the complete log-likelihood as
\[
\ell^*(\b\eta)=\ell_1^*(\b\eta_1)+\ell_2^*(\b\eta_2),
\]
with
\begin{eqnarray}
\ell_1^*(\b\eta_1)&=&\sum_c m_c\log\pi_c,\label{eq:comp_lk1}\\
\ell_2^*(\b\eta_2)&=&\sum_c\sum_j\b m_{cj}\tr\log\b\la_{cj},
\label{eq:comp_lk2}
\end{eqnarray}
where $m_c=\sum_{\bl x}m_{c,\bl x}$ is the number of subjects in latent class $c$
and $\b m_{cj}$ is the column vector with elements $\sum_{\bl x}I(x_j=x)m_{c,\bl x}$,
$x=1,\ldots,l_j-1$, with $I(\cdot)$ denoting the indicator function.

The EM algorithm alternates the following
two steps until convergence:
\begin{description}
\item[E-step:] compute the conditional expected value of $\ell^*(\b\eta)$
given the observed data and the current value of the parameters;
\item[M-step:] maximize the above expected value with respect to $\b\eta$,
so that this parameter vector results updated.

\end{description}

The E-step consists of computing, for every $c$ and $\b x$, the expected value
of $m_{c,\bl x}$ given $n_{\bl x}$ as follows
\[
\hat{m}_{c,\bl x} = n_{\bl x} \frac{p(\b x|c)\pi_c}{\sum_h p(\b
x|h)\pi_h}
\]
and then substituting these expected frequencies in (\ref{eq:comp_lk}). On the
basis of $\hat{m}_{c,\bl x}$ we can obtain the expected frequencies $\hat{m}_c$
and $\hat{\b m}_{cj}$ which, once substituted in (\ref{eq:comp_lk1}) and
(\ref{eq:comp_lk2}), allow us to obtain the expected values of $\ell_1^*(\b\eta_1)$
and $\ell_2^*(\b\eta_2)$, denoted by $\hat{\ell}_1^*(\b\eta_1)$ and
$\hat{\ell}_2^*(\b\eta_2)$, respectively.

At the M-step, the function obtained as described above is maximized with respect
to $\b\eta$ as follows. First of all, regarding the parameters in
$\b\eta_1$ we have
an explicit solution given by
\[
\pi_c= \frac{\hat{m}_c}{n},\quad c=2,\ldots,k, 
\]
which corresponds to the maximum of $\hat{\ell}^*_1(\b\eta_1)$.
To update the other parameters, we maximize $\hat{\ell}^*_2(\b\eta_2)$
by a Fisher-scoring algorithm that we illustrate in the following.

The Fisher-scoring algorithm alternates a step in which the parameter vector $\b\ga$
is updated with a step in which the parameter vector $\b\phi$ is updated.
The first step consists of adding to the current value of each
free $\ga_j$ the ratio $s_{2j}^*/f_{2j}^*$,
where $s_{2j}^*$ denotes the score for $\hat{\ell}^*_2(\b\eta_2)$
with respect to $\ga_j$ and $f_{2j}^*$ denotes the corresponding
information computed at the current value of the parameters. These have
the following expressions:
\begin{eqnarray*}
s_{2j}^* &=& \sum_c\sum_j(\b Z_{cj}\b\phi)\tr
\b R_{cj}\tr(\hat{\b m}_{cj}-\hat{m}_c\b\la_{cj}),\\
f_{2j}^* &=& \sum_c\hat{m}_c\sum_j(\b Z_{cj}\b\phi)\tr
\b R_{cj}\tr[\diag(\b\la_{cj})-\b\la_{cj}\b\la_{cj}\tr]\b R_{cj}(\b Z_{cj}\b\phi),
\end{eqnarray*}
where $\b R_{cj}$ is the derivative matrix of the canonical parameter vector
for $\b\la_{cj}$ with respect to the vector of logits in (\ref{eq:logits});
see \cite{colo:forc:01}.
Then, the parameter vector $\b\phi$ is updated by adding the quantity
$(\b F_2^*)^{-1}\b s_2^*$, where $\b s_2^*$ is the score vector for
$\hat{\ell}^*_2(\b\eta_2)$ with respect to $\b\phi$ and $\b F_2^*$ denotes the
corresponding information computed at the current parameter value, which
have the following expressions:
\begin{eqnarray*}
\b s_2^* &=& \sum_c\sum_j\ga_j\b Z_{cj}\tr
\b R_{cj}\tr(\hat{\b m}_{cj}-\hat{m}_c\b\la_{cj}),\\
\b F_2^* &=& \sum_c\hat{m}_c\sum_j\ga_j^2\b Z_{cj}\tr
\b R_{cj}\tr[\diag(\b\la_{cj})-\b\la_{cj}\b\la_{cj}\tr]\b R_{cj}\b Z_{cj}.
\end{eqnarray*}

As usual, we suggest to initialize the EM algorithm by a deterministic
rule and by a multi-start strategy based on random starting values which
are suitable generated. In this way we can deal with the multimodality of
the model likelihood.
\subsection{Model selection} \label{selec}
The formulation of a specific model in the class of multidimensional LC IRT models for
ordinal responses univocally depends on: ({\em i}) the number of latent classes ($k$); 
({\em ii}) the adopted parameterization in terms of link function $g_x(\cdot)$ and
constraints on the item parameters
$\ga_j$ and $\beta_{jx}$, and ({\em iii}) the number ($s$) of latent dimensions and the
corresponding allocation of items within each dimension ($\de_{jd}$, $j=1,\ldots,r$,
$d=1,\ldots,s$).
Thus, the model selection implies the adoption of a number of choices, for each of the 
previously mentioned aspects, by using suitable criteria. In the following, we mainly 
refer on the likelihood ratio (LR) test and 
on the Bayesian  \citep{sch:78} information criterion (BIC). Firstly, we briefly
recall these methods; then, we illustrate in detail the 
suggested model selection procedure.
\subsubsection{Criteria for model selection}
As it is well known, given a certain hypothesis denoted by $H_0$,
the LR test is based on the statistic
\[
D = -2 (\hat{\ell}_0  - \hat{\ell}),
\]
where $\hat{\ell}_0$ and  $\hat{\ell}$  denote the maximum of log-
likelihood of the reduced model which incorporates $H_0$ and under the general model,
respectively. Under this hypothesis, and provided that suitable regularity conditions hold,
LR statistics is asymptotically distributed as a $\chi_q^2$,
where $q$ is given by the difference in the number of parameters between the two nested
models being compared. An asymptotically equivalent alternative to LR test,
is the Wald test, which, however, requires to compute the information matrix
of the model.

Differently from the LR (and the Wald) statistics, information criteria do not provide
 neither
a test of a model in the usual sense of testing a null hypothesis nor information about
the way a model fits the data in absolute terms. However, they offer a relative measure
of lost information when a given model is used to describe observed data. Besides, they
are particularly useful to select among two or more general models, especially 
non-nested models, that cannot be compared by means of LR or Wald tests.

Different types of information criteria have been proposed in the statistical 
literature, 
and among them we prefer the Bayesian Information Crierion \citep[BIC, ][]{sch:78},
which is based on introducing  a penalty term in the model to take into account the
number of parameters. More precisely, this criterion is based on the index:
\[
BIC = -2\hat{\ell} + \log(n)\#{\rm par},
\]
where $\hat{\ell}$ is the maximum value of the log-likelihood of the model of interest,
and $\#{\rm par}$ is the number of free parameters defined in Table \ref{tab:num_par}.
The smallest the BIC index is, the better is the model fitting. Therefore, among a set
of competing models, we choose that with the minimum BIC value.

BIC has to be preferred to other information
criteria, because it satisfies some nice properties. Mainly, under certain regularity
conditions it is asymptotically consistent \citep{keribin:2000}. Moreover,
since it applies a larger penalty for additional parameters (for reasonable
sample sizes) in comparison with other criteria, BIC tends to select more parsimonious
models.
\subsubsection{Model selection procedure}\label{sec:modelsel}
As stressed at the beginning of this section, the specification of a multidimensional LC
IRT model for ordinal items implies 
a number of choices. A model
selection procedure is here proposed which is based on the following sequence of ordered steps:
\begin{enumerate}
\item selection of the optimal number $k$ of latent classes;
\item selection of the type of link function;
\item selection of the number of latent dimensions and item allocation
within each dimension;
\item selection of constraints on the item discriminating and difficulty parameters.
\end{enumerate}
These steps are described in more detail in the following.
\begin{enumerate}
\item \emph{Selection of the number of latent classes.} To detect the optimal number $k$
of latent classes, it is useful to proceed by comparing models that differ only in the
number of latent classes, all other features being equal. More precisely, we suggest to
adopt the standard LC model \citep{good:74}, characterized by one dimension for each
item. In this way, no choice on the link function and the item parameterization is 
requested; also, any restrictive assumptions on item dimensionality is avoided.

To compare LC models we rely on BIC, as it is not feasible to compare LC models with 
different number of latent classes through an LR test statistic. In particular,
we fit the LC model with increasing $k$ values; then, the value just before
the first increasing BIC index is taken as optimal number of latent classes.

A crucial problem with LC models is represented by the multimodality of the
likelihood function.
To avoid that the choice of $k$ falls in correspondence
of a local - rather than global - maximum point, we suggest to repeat the estimation
process by randomly varying the starting values of the model parameters. Then, for each
possible value of $k$, we select the highest obtained log-likelihood value and, 
consequently, the smallest estimated BIC value.
\item\emph{Selection of the logit link function.} As described in Section
\ref{prelim}, it is possible to choose among three different types of logit:
local logits, global logits, and continuation
ratio logits. In particular, we perform the comparison between models on the basis of 
the mentioned logit functions and adopting BIC, which is here preferred to 
the LR test statistic as the latter cannot be validly used when models are not nested. 
Besides, when comparing the models, we choose the number of latent classes as selected 
in the previous step and we adopt the same multidimensional latent structure, 
that is with one dimension for each item. As concerns the item parameterization,
we suggest to choose the most general one, which is
based on both free item discriminating parameters and on free item difficulties parameters.

Obviously, since it can happen that no relevant difference in
the goodness of fit of the competing models comes out (i.e., BIC index assumes very 
similar values), the choice of the type of logit should also take into account
the different interpretations behind the three types of logits;
see also \cite{oli:94} and \cite{sam:96}.
\item \emph{Selection of dimensions.} Detection of latent traits
is of main interest when estimating multidimensional IRT models.
Several authors have dealt with testing unidimensionality in connection with Rasch type
models. One of the main contributions is due to \cite{mar:73},
who developed an LR test for
the unidimensionality assumption against the alternative that the items consist of two
subsets, defined in advance, each measuring one latent trait. This test has been
generalized through a conditional non-parametric approach by \cite{chr:02} to the case
of polytomous items and to cases with more than two dimensions.

To the aim of detecting latent traits in a more general context than that of Rasch
type models, the LR statistic may be used to test the
unidimensionality of a set of items against a specific multidimensional alternative,
being the null hypothesis specialized as
\[
H_0: \theta_{d_1c}=  a_{d_1d_2}+b_{d_1d_2}\theta_{d_2c},
\quad\forall d_1 \neq d_2 = 1, \ldots, s,
\]
for two constants $a_{d_1d_2}$ and $b_{d_1d_2}$,
where the second is equal to 1 if the parametrization
based on the constant discrimination indices is assumed.

For instance, in the case of two dimensions, we compare a model in which these
dimensions are collapsed (unidimensionality assumption) with a model in which they are
kept distinct (bidimensionality assumption), all other elements being equal, in
accordance with the results of the previous steps.
On the basis of this principle, \cite{bart:07} proposes a model-based hierarchical
clustering procedure that can also be applied for the extended models here proposed to
take into account ordinal items and that allows us to detect groups of items that
measure the same latent trait. 
\item \emph{Choice of the item discriminating and difficulty parameterization.}
This step consists of the choice of the possible constraints on the
discriminating and difficulty parameters. Four
different types of model may be defined by combining free or constrained  $\gamma_j$
parameters with free or constrained $\beta_{jx}$ parameters. Once the other elements of
the model have been defined through the previous steps,
we may perform the comparison 
among the four models on the basis of the LR (or Wald) test. Indeed, the
null hypothesis $H_0$ we are testing when we compare a
model with free $\gamma_j$ with a model with constrained $\gamma_j$ is the same as that
expressed in (\ref{eq:same_discrimination}). Similarly,
by decomposing the item difficulty parameters as sum of two components,
that is $\beta_{jx}=\beta_j+\tau_{jx}$, where
$\tau_{jx}$ is referred to item $j$ and category $x$, and maintaining
the same assumption about discriminating parameters,
we easily realize that hypothesis (\ref{eq:rating_scale})
is equivalent to
\[
H_0: \tau_{jx}= \tau_j,\quad j=1,\ldots,r,\: x=1,\ldots,l-1,
\]
which can be still tested by an LR statistic.
\end{enumerate}
\section{Application to measurement of anxiety and depression} \label{applic}
The data used to illustrate the proposed class of
polytomous LC IRT models concerns a sample of 201 oncological
Italian patients who were asked to fill in questionnaires about their health and
perceived quality of life. Here we are interested in anxiety and
depression, as assessed by the ``Hospital Anxiety and Depression Scale'' (HADS)
developed by \cite{zig:sna:1983}.
The questionnaire is composed by $14$ polytomous items equally divided between the two
dimensions:

\begin{enumerate}
\item anxiety (7 items: 2, 6, 7, 8, 10, 11, 12);
\item depression (7 items: 1, 3, 4, 5, 9, 13, 14).
\end{enumerate}

Apparently, within this context of study, the assumption of unidimensionality might be
not realistic. Thus, the adoption of the proposed class of models, rather than a 
unidimensional
IRT model, appears more suitable and well more convenient, as it allows to detect
homogeneous classes of individuals who have similar latent characteristics, so that
patients in the same class will receive the same clinical treatment.

All items of the HADS questionnaire have four response categories: the minimum value 0
corresponds to
a low level of anxiety or depression, whereas the maximum value 3 corresponds to a
high level of anxiety or depression. Table \ref{tab:descr} shows the
distribution of item responses among the four categories, distinguishing between the two
supposed dimensions.

\begin{table}[!ht]\centering
\vspace*{0.5cm}
{\small
\begin{tabular}{l|cccc|c}
\hline\hline
   &     \multicolumn{4}{c|}{Response category}      &   \\
\hline
Item	&	0	&	1	&	2	&	3	&	Total 	\\
\hline
2	&	35.3	&	52.7	&	8.0	&	4.0	&	100.0	\\
6	&	39.8	&	46.3	&	10.0	&	4.0	&	100.0	\\
7	&	46.3	&	22.4	&	21.9	&	9.5	&	100.0	\\
8	&	19.4	&	49.3	&	24.9	&	6.5	&	100.0	\\
10	&	7.0	&	40.8	&	44.3	&	8.0	&	100.0	\\
11	&	30.8	&	49.8	&	11.4	&	8.0	&	100.0	\\
12	&	34.3	&	46.3	&	14.9	&	4.5	&	100.0	\\
\hline\hline
Anxiety	&	30.4	&	43.9	&	19.3	&	6.3	&	100.0	\\
\hline
1	&	43.8	&	32.8	&	16.4	&	7.0	&	100.0	\\
3	&	56.7	&	29.9	&	9.0	&	4.5	&	100.0	\\
4	&	31.8	&	54.7	&	11.9	&	1.5	&	100.0	\\
5	&	46.3	&	38.8	&	13.4	&	1.5	&	100.0	\\
9	&	9.0	&	27.9	&	55.2	&	8.0	&	100.0	\\
13	&	42.3	&	42.3	&	11.4	&	4.0	&	100.0	\\
14	&	30.8	&	37.3	&	28.9	&	3.0	&	100.0	\\
\hline\hline
Depression	&	37.2	&	37.7	&	20.9	&	4.2	&	100.0	\\
\hline
\end{tabular}}
\caption{\em Distribution of HADS item responses (row percentage frequencies).}
\label{tab:descr}\vspace*{0.5cm}
\end{table}

Altogether, responses are mainly concentrated in categories 0 and 1 both for anxiety and
depression, whereas category 3, that denotes high levels of psychopathological
disturbs, is selected less than $10\%$ of the times for each item.  By summing item
responses, it is possible to obtain, for each patient, a score indicating a raw measure
of anxiety and depression: the closer the raw score is to the minimum value 0, the lower
the level of anxiety or depression is, and viceversa. The mean raw score observed for
the entire sample is very similar through the two dimensions, being  7.11 for anxiety
and 7.17 for depression (standard deviation is equal to 4.15 and 4.16, respectively).
Correlation between scores on anxiety and scores on depression is very high; it is
equal to 0.98.

To proceed to the model selection, the four ordered steps suggested in Section
\ref{sec:modelsel} are followed.
We recall that the first step consists of detecting the optimal number $\hat{k}$ of
latent classes. To this aim, the standard LC model is employed and a comparison among
models which differ by the number of latent classes is performed for $k = 1, 2, 3,
4$. The results of this preliminary fitting are reported in Table \ref{table5}, where,
to avoid the multimodality problem, results are referred both to deterministic and to
random starting values.

\begin{table}[!ht]\centering
\vspace*{0.5cm}
{\small
\begin{tabular}{l|ccc|ccc}
\hline\hline
 &      \multicolumn{3}{c|}{Deterministic start}      & \multicolumn3c{Random start} \\
\hline
  $k$           &  \multicolumn{1}{c}{$\hat{\ell}$}        &  \ {$\#{\rm par}$}  & 
\multicolumn{1}{c|}{BIC}  & \multicolumn1c{$\hat{\ell}$(max)} &  \ {$\#{\rm par}$}  & 
\multicolumn1c{BIC(min)}       \\
\hline
1	& -3153.151 & 42 &  6529.040				&	-3153.151 & 42  & 
6529.040	  		\\
2	& -2814.635	& 85 &  6080.051	               & 	-2814.635 & 85 &  
6080.051	   	\\
3	& -2677.822 & 128 & \bf 6034.468			&	-2674.484  & 128 & 
\bf 6027.791 	 	\\
4	& -2645.435	 & 171 & 6197.736	       		&	-2608.570 & 171  & 
6104.805 			\\
\hline
\end{tabular}}
\caption{\em Standard LC models: log-likelihood ($\hat{\ell}$), number of parameters, 
and BIC values  for $k = 1,
\ldots, 4$ latent classes; in boldface is the smallest BIC
value, selected with deterministic and random starts.}
\label{table5}\vspace*{0.5cm}
\end{table}

On the basis of the adopted selection criterium, 
we choose $\hat{k} = 3$ as optimal number of 
latent classes as, in correspondence of this number of latent classes, the smallest 
estimated BIC value is observed, both with a deterministic and random initialization
of the EM algorithm.

As regard to the second step and the choice of the best logit link function,  a 
comparison between a graded response type model and a partial credit type model is 
carried out by assuming $\hat{k} = 3$ latent classes, free item discriminating and 
difficulties parameters, a completely general multidimensional structure for the data 
(i.e., $r$ dimensions, one for each item), and basing the comparison on the BIC index.
Note that the continuation ratio logit link function is not suitable in this context, 
because the item response process does not consist of a sequence of successive steps. 
Table \ref{table6} shows that a global logit link  has to be preferred to a local logit 
link. Also, it can be observed that a graded response type model has a better fit than 
the standard LC model, as the BIC value observed for the former is smaller than that 
detected for the latter (see Table \ref{table5}).

\begin{table}[ht!]\centering\vspace*{0.5cm}
\small
\begin{tabular}{l|cc}
\hline\hline
 &  Global logit     &        Local logit         \\
\hline
$\hat{\ell}$	&	-2726.348	&	-2741.321	\\
{$\#{\rm par}$} & 72 & 72 \\
BIC	&	\bf 5834.534	&	5864.479\\
\hline
\end{tabular}
\caption{\em Graded response and partial credit type models with $\hat{k}=3$: 
log-likelihood ($\hat{\ell}$), number of parameters, and BIC values; in boldface is the 
smallest BIC value.}
\label{table6}\vspace*{0.5cm}
\end{table}

Once we have chosen the global logit as the best link function, we carry on with the 
test of unidimensionality. 
An LR test is used to compare models which differ on account of their dimensional 
structure, all other elements being equal (i.e., free item discriminating and difficulty 
parameters), that is (\textit{i}) a graded response model with $r$-dimensional 
structure, (\textit{ii}) a graded response model with bidimensional structure (i.e., 
anxiety and depression), and (\textit{iii}) a graded response model with unidimensional 
structure (i.e., all the items belong to the same dimension). For the sake of 
completeness, log-likelihood and BIC values are also provided for each model
considered. On account of both BIC and the LR
test, the hypothesis of unidimensionality may be accepted (see Table \ref{table7}). This
result is coherent with a similar analysis performed on the same data by Bacci and Bartolucci (to appear),
where item responses were dichotomized and a Rasch parameterization was adopted.

\begin{table}[!ht]\centering\vspace*{0.5cm}
\small\begin{tabular}{l|ccccc}
\hline\hline
Model               &    $\hat{\ell}$ & {$\#{\rm par}$}    &    BIC     &     Deviance   &   $p$-value   \\
\hline
$r$-dimensional     &    -2726.348 & 72 & 5834.534   &      --  &       --   \\
bidimensional       &    -2731.249 & 60 & 5780.696   &     $\:$9.802    &       0.633  \\
unidimensional &     -2731.894 & 59	&    \bf 5776.682   &        $\:$1.290  &  0.256   \\
\hline
\end{tabular}
\caption{\em $r$-dimensional, bidimensional, and unidimensional graded response models with $\hat{k}=3$: log-likelihood, number of parameters, BIC value,
and LR test results (deviance and $p$-value); in boldface the smallest BIC value.}
\label{table7}\vspace*{0.5cm}
\end{table}

As previously outlined, the choice of the number of parameters per item depends on both
the presence of a constant/non-constant discriminating index ($\gamma_j$), and of a
constant/non-constant threshold difficulty parameter ($\beta_{jx}$), for each item. In
our application, this implies a comparison among four models, in accordance with the
classification adopted in Table \ref{tab:classif}.
The parameterization is chosen on account of the unidimensional data structure and the
previously selected global logit link function. Besides, because the compared models are
nested, the parameterization is selected on the basis of an LR test. Again, for the sake 
of completeness, log-likelihood and BIC
values are also provided for each model considered.

The analyses show (Table~\ref{table8}) that between GRM and RS-GRM, GRM has to be
preferred to RS-GRM, while between models GRM and 1P-GRM, the latter has to be
preferred. Besides, as model 1P-GRM has a better fit than model 1P-RS-GRM, then 1P-GRM
has to be preferred model among the four considered, that is the graded response type
model with free $\beta_{jx}$ parameters 
and constant $\gamma_j$ parameters. Such a result is achieved by
taking into account both the BIC criterium and the LR test.

\begin{table}[!ht]\centering\vspace*{0.5cm}
\small\begin{tabular}{l|cccclc}
\hline\hline
Model         &       $\hat{\ell}$    & {$\#{\rm par}$}   &    BIC     &     
\multicolumn2c{Deviance}    &
$p$-value   \\
\hline
GRM       &      -2731.894        &     59 &    5776.682       &        
\multicolumn2c{--} &       --   \\
RS-GRM  &        -2795.570   &    33  &   5766.149    &     127.353 & (vs GRM) & 0.000 \\
1P-GRM &       -2741.285     &  46 &       \bf 5726.521      &    \;\;18.782 & (vs GRM) 
& 0.130 \\
1P-RS-GRM &  -2844.518     & 20 &    5795.102     &  206.467 & (vs 1P-GRM) & 0.000    \\
\hline
\end{tabular}
\caption{\em Item parameters selection: log-likelihood, number of parameters, BIC values,
and LR test results
(deviance and $p$-value) between nested graded response models with $\hat{k}=3$ and $s=1$; in 
boldface the smallest BIC value.}
\label{table8}\vspace*{0.5cm}
\end{table}

As the sequence of the previously described steps may be considered partly arguable, it
can be also shown that the same results - in terms of link function, item
parameterization and dimensionality choice - would have been obtained if each of such
models were compared at once accounting for log-likelihood and BIC values as selection
criteria. Indeed, Table \ref{table9} shows that the smallest BIC value is observed when
selecting: (\textit{i}) a global logit link function; (\textit{ii}) constrained
$\gamma_j$ parameters and
free $\beta_{jx}$ parameters, that is, a 1P-GRM model; and (\textit{iii}) assuming a
unidimensional structure for the data. 

\begin{table}[!ht]\centering
\vspace*{0.5cm}
{\small
\begin{tabular}{c|c|c|cc|cc}
\hline\hline
Dimensionality & \multicolumn{2}{c|}{Item parameters} &     \multicolumn{2}{c|}{Global
logit}      & \multicolumn2c{Local logit}   \\
\hline
           & \multicolumn1c{$\gamma_j$}
           & \multicolumn{1}{c|}{$\beta_{jx}$}
           &       \multicolumn1c{$\hat{\ell}$}   &
           \multicolumn{1}{c|}{BIC}        &
           \multicolumn1c{$\hat{\ell}$}    &
           \multicolumn1c{BIC}     \\
\hline
$r$-dimensional 	&	free/constr.	&	free	&	-2726.347   &  5834.534 & -2741.321   
& 5864.479  	\\	            
	                        &	free/constr.	&	constrained	& -2815.568	&  5875.088 
	                        & -2836.766   &  5917.484  \\
	                        \hline
bidimensional 	&	free	&	free	&	-2731,249	&	5780,696	&	-2749,839	
&	 5817,877	\\
	&	constrained	&	free	&	-2740,658	&	5735,875	&	-2764,787	&	
	5784,132	 \\
	&	free	&	constrained	&	-2798,959	&	5778,230	&	-2835,611	&	
	5851,534	 \\
	&	constrained	&	constrained	&	-2843,227	&	5803,127	&	-2869,223	&	
	5855,120	\\
\hline
unidimensional 	&	free	&	free	&	-2731,894	&	5776,682	&	-2750,214
	&	 5813,323	\\
	&	constrained	&	free	&	-2741,285	&	\bf 5726,521	&	-2765,129	&	
	5774,211	\\
	&	free	&	constrained	&	-2795,570	&	5766,149	&	-2833,179	&	
	5841,366	 \\
	&	constrained	&	constrained	&	-2844,518	&	5795,102	&	-2870,178	&	
	5846,422	\\
\hline
\end{tabular}}
\caption{\em Log-likelihood and BIC values for the global and local logit link
functions, taking into account the dimensional structure ($r$-dimensional/bidimensional/
unidimensional)
and the item parameters (depending on whether they are free/constraint); in boldface is
the smallest BIC value.}
\label{table9}\vspace*{0.5cm}
\end{table}

The estimates of support points $\hat{\xi}_c$ and probabilities $\hat{\pi}_c$,
$c=1,2,3$, under the selected unidimensional 1P-GRM model are shown in Table
\ref{table10}. On the basis of these results, we conclude
that patients who suffer from
psychopatological disturbs are mostly represented in the first two classes, whereas only
the $16.7\%$ of the subjects belong to the third class. Furthermore, patients belonging
to class 1 present the least severe conditions, whereas
patients in class 3 present the worst conditions.

\begin{table}[!ht]\centering\vspace*{0.5cm}
\small\begin{tabular}{l|ccc}
\hline\hline
      &    \multicolumn3c{Latent class $c$} \\
\hline
Dimension       &   \multicolumn1c{1}   &   \multicolumn1c{2}   &   \multicolumn1c{3} \\
\hline
Psychopatological disturbs &   -0.776      &  1.183      &   3.418    \\
Probability &  0.342   &   0.491   &   0.167    \\
\hline
\end{tabular}
\caption{\em Estimated support points $\hat{\xi}_c$ and probabilities
$\hat{\pi}_c$ of latent classes for the unidimensional 1P-GRM.} \label{table10}\vspace*{0.5cm}
\end{table}
\newpage
\section{Concluding remarks}\label{sec:conclusion}
In this article, we extend the class of multidimensional latent class (LC) Item Response
Theory (IRT) models \citep{bart:07} for dichotomously-scored 
items to the case of ordinal
polytomously-scored items. The proposed models are
formulated in a general way, so that
several different parameterizations may be adopted for the distribution of the response
variable, conditioned to the vector of latent traits. The classification criteria we use
are based on three main elements: the type of link function, which may be 
based on global, local, or continuation ratio logits,
the type of constraints on item discriminating parameters,
that may be completely free or kept all equal to one, and the type of constraints on
item  difficulty parameters, that may be formulated so that each item has different
distances between consecutive response categories or in a more parsimonious way, where
the distance between difficulty levels from category to category within each item is the
same across all items. According to the way these criteria are combined, twelve possible
parameterizations result, some of which are well-known in the psychometrical literature,
such as those referred to the Graded Response Model \citep{sam:69}, the Partial Credit
Model \citep{mast:82}, and the Rating Scale Model \citep{andr:78}.

The proposed class of models is more flexible in comparison with traditional formulations of
IRT models, often based on restrictive assumptions, such as unidimensionality and
normality of latent trait. In particular, the assumption of multidimensionality allows
us to take more than one latent trait into account at the same time and to study the
correlation between latent traits. Moreover, in the proposed class of models, no
specific assumption about the distribution of latent traits is necessary, since a latent
class approach is adopted, in which the latent traits are represented by a random vector
with a discrete distribution common to all subjects. In this way, subjects with similar
latent traits are assigned to the same latent class, so as to detect homogeneous
subpopulations of subjects. 
 Moreover, the latent class approach presents a notable simplification from the computational point of view with respect to the case of continuous latent traits, where the marginal likelihood is characterized by a multidimensional integral difficult to treat.  

In order to make inference on the proposed model, we show how the
log-likelihood may be efficiently maximized by the EM algorithm. We 
also propose a model selection procedure to choose the
different features that contribute to define a specific multidimensional LC IRT model.
In general, comparisons between different parameterizations are based on information
criteria, in particular we rely on the Bayesian Information Criterion \citep{sch:78}
or on likelihood ratio (or Wald) test, being this last tool
useful in presence of nested models. First of all, we suggest to verify  the
reasonableness of the discreteness assumption by selecting the number of latent classes.
In order to obtain a more parsimonious model, this first phase of the model selection
should be performed with reference to the standard LC model. Then, given the selected
number of latent classes and the most general parameterization about items and
dimensionality, the choice among global, local or continuation ratio logit link
functions may be performed, so that a graded response or a partial credit or a
sequential model is selected. This phase should also take into account the
interpretability of the type of logit with reference to the specific application
problem. The next phase consists of choosing the number of latent dimensions and the 
allocation of items within each
dimension. This phase may be more or less complex depending on a
priori information about the dimensionality structure of the questionnaire. Finally, 
possible constraints on the item
discriminating and difficulty parameters are selected, by comparing nested models
that are equal as concerns all the other elements. 

The class of multidimensional LC IRT models for ordinal items and the proposed model
selection procedure are illustrated through an application to a dataset, which 
concerns the measurement of psychopathological disturbs 
(i.e., anxiety and depression) in oncological
patients by using the Anxiety and Depression Scale of \cite{zig:sna:1983}. The results
show that subjects can be classified in three latent classes, and the item responses can
be explained by a graded response type model with items having the same discriminating
power and different distances between consecutive response categories. The
bidimensionality assumption is rejected in favor of unidimensionality, so that all
items of the questionnaire measure the same latent  psychopathological disturb.

{\small
\bibliography{biblio}
\bibliographystyle{apalike}}

\end{document}